# DYNAMIC SCHEDULING OF SKIPPABLE PERIODIC TASKS WITH ENERGY EFFICIENCY IN WEAKLY HARD REAL-TIME SYSTEM

Santhi Baskaran[1] and P. Thambidurai[2]

[1]Department of Information Technology, Pondicherry Engineering College,
Puducherry, India
santhibaskaran@pec.edu

[2]Department of Computer Science and Engineering, Pondicherry Engineering College,
Puducherry, India
ptdurai@pec.edu

## ABSTRACT

*Energy consumption is a critical design issue in real-time systems, especially in battery- operated systems. Maintaining high performance, while extending the battery life between charges is an interesting challenge for system designers. Dynamic Voltage Scaling (DVS) allows a processor to dynamically change speed and voltage at run time, thereby saving energy by spreading run cycles into idle time. Knowing when to use full power and when not, requires the cooperation of the operating system scheduler. Usually, higher processor voltage and frequency leads to higher system throughput while energy reduction can be obtained using lower voltage and frequency. Instead of lowering processor voltage and frequency as much as possible, energy efficient real-time scheduling adjusts voltage and frequency according to some optimization criteria, such as low energy consumption or high throughput, while it meets the timing constraints of the real-time tasks. As the quantity and functional complexity of battery powered portable devices continues to raise, energy efficient design of such devices has become increasingly important. Many real-time scheduling algorithms have been developed recently to reduce energy consumption in the portable devices that use DVS capable processors. Extensive power aware scheduling techniques have been published for energy reduction, but most of them have been focused solely on reducing the processor energy consumption. While the processor is one of the major power hungry units in the system, other peripherals such as network interface card, memory banks, disks also consume significant amount of power. Dynamic Power Down (DPD) technique is used to reduce energy consumption by shutting down the processing unit and peripheral devices, when the system is idle. Three algorithms namely Red Tasks Only (RTO), Blue When Possible (BWP) and Red as Late as Possible (RLP) are proposed in the literature to schedule the real-time tasks in Weakly-hard real-time systems. This paper proposes optimal slack management algorithms to make the above existing weakly hard real-time scheduling algorithms energy efficient using DVS and DPD techniques.*

## 1. INTRODUCTION

Battery powered portable real-time systems have been widely used in many applications. As the quantity and the functional complexity of battery powered portable devices continues to raise, energy efficient design of such devices has become increasingly important. Also, these real-time systems have to concurrently perform a multitude of complex tasks under stringent time constraints. Thus, minimizing power consumption and extending battery lifespan while guaranteeing the timing constraints has become a critical aspect in designing such systems. The interest in real-time systems has been growing steadily since more industrial systems rely on International Journal of Computer Science & Information Technology (IJCSIT), Vol 2, No 6, December 2010computer based operations. Therefore, the critical applications are being done by the computer in real-time environment must produce desired result at the correct time. The result (correct output) not obtained in correct time may be disastrous. As per the definition, the output of real-time systems not only depends on the correctness of the result but also the time when the result is produced.

Based on the functional criticality of jobs, usefulness of late results and deterministic or probabilistic nature of the constraints, the real time systems are classified as, *Hard real-time system* in which consequences of not executing a task before its dead line catastrophic or fatal, *Soft real-time system* in which the utility of results produced by a task decreases over time after deadline expires and *Firm* or *Weakly hard real-time system* in which the result produced by a task ceases to be useful as soon as the deadline expires but the consequences of not meeting the deadline are not very severe [1]. Typical illustrating examples of systems with weakly-hard real-time requirements are multimedia systems in which it is not necessary to meet all the task deadlines as long as the deadline violations are adequately spaced [2].

Computations occurring in a real-time system that have timing constraints are called real-time tasks. A real-time application usually consists of a set of cooperating tasks activated at regular intervals and/or on particular events. Tasks in real-time system are of two types, periodic tasks and aperiodic tasks [1]. Periodic tasks are time driven and recur at regular intervals called the period. Aperiodic tasks are event driven and activated only when certain events occur. The necessary condition is that real-time tasks must be completed before their deadlines for a system to be successful.

Weakly hard real-time systems research is motivated by the observation that for many real-time applications (which are periodic in nature) some deadline misses are acceptable as long as they are spaced distantly/evenly. Examples for such applications include multimedia processing, real-time communication and embedded control applications. There have been some previous approaches to the specification and design of real-time systems that tolerate occasional losses of deadlines. Hamdaoui and Ramanathan introduced the idea of (m, k)-firm deadlines [3] to model tasks that have to meet m deadlines every k consecutive instances. If this constraint is violated in any time window, the system is said to exhibit a dynamic failure (implying possible degradation in system performance or quality-of-service). The Skip-Over model was introduced by Koren and Shasha [4] with the notion of skip factor. In this model, a task's tolerance to deadline misses is characterized by the skip factor *s*: in any *s* consecutive instances of the task at most one can miss its deadline. It is a particular case of the (m, k)-firm model. They reduce the overload by skipping some task instances, thus exploiting skips to increase the feasible periodic load. In the Dynamic Window Constrained Scheduling (DWCS) model motivated by the real-time packet scheduling applications, a given task needs to complete at least *m* instances before their deadlines in every non-overlapping window of *k* instances [5], [6], [7].

In real-time systems, the systems must schedule the tasks by deadlines and there is no benefit in finishing the computation early. Making computations energy efficient in the systems, the battery lifetime can be increased. In order to make them energy efficient, in the scheduling, the execution time of the tasks can be extended up to the deadline for each task set. This is possible through dynamic voltage scaling (DVS) technique.

In this paper, we address the problem of the dynamic scheduling of periodic task sets with skip constraints. In this context, the objectives of a scheduling algorithm are to maximize the effective Quality of Service (QoS) of periodic tasks defined as the number of task instances which complete before their deadline and to minimize the energy consumption of tasks.

This paper is organized in the following way. The processor, energy and resource task models are described in Section 2. Existing scheduling algorithms (without energy efficiency) for weakly-hard real-time systems are explained in Section 3. Energy-efficiency technique

101

proposed to the existing algorithms is described in Section 4. Section 5 discusses the simulation and analysis of results. Finally Section 6 concludes this paper with future work.

## 2. SYSTEM MODELS

In this section, we briefly discuss the processor, energy and task models that we have used in our work.

### 2.1. Processor Model

The target platform of this work is a single processor system whose only power source is a battery. We assume that the system has DVS capability, where the processor speed (frequency) and supply voltage can be dynamically adjusted. We further assume the processor can exist in two modes: *execution mode* and *stand-by mode.* In the stand-by mode, the processor does not execute any tasks, and consumes only the stand-by power. The CPU switches to stand-by mode if it is idle. In the execution mode, the CPU speed can vary between a lower bound $S_{min}$ and an upper bound $S_{max}$. In this case, the power consumed is a function of the CPU speed/frequency. In any time interval $[\tau_1, \tau_2]$, the total energy consumption is the integral of power consumption function, which includes the stand-by and dynamic power consumption components. We also assume that time and energy overheads due to CPU speed changes are negligible. We adopt an *inter-task DVS* model; that is, we assume that the CPU speed can be changed only at task completion or pre-emption points, following [8], [9].

### 2.2. Energy Model

The DVS technique reduces the dynamic power dissipation by dynamically scaling the supply voltage and the clock frequency of processors. The relationship between power dissipation $P_d$, supply voltage $V_{dd}$, and frequency $f$ is represented by

$$P_d = C_{ef} \text{ X } V^2_{dd} \text{ X } f \quad \text{and}$$
$$f = k \text{ X } (V_{dd} - V_t^2)/V_{dd},$$

where $C_{ef}$ is the switched capacitance, k is the constant of circuit, and $V_t$ is the threshold voltage [10]. The energy consumed to execute task $T_i$, $E_i$, is expressed by $E_i = C_{ef} X V^2_{dd} X \varepsilon_i$, where $\varepsilon_i$ is the number of cycles to execute the task. The supply voltage can be reduced by decreasing the processor speed. It also reduces energy consumption of task. Here we use the task's execution time at the maximum supply voltage during assignment to guarantee deadline constraints.

### 2.3. Task Model

We consider a set of *n* independent periodic real-time tasks $\Gamma = \{T_1, T_2, …, T_n\}$. Each task $T_i = (p_i, c_i, a_i, sf_i)$ is characterized by four parameters: the period $p_i$, worst-case execution time $c_i$ which is the upper bound on the computation time of $T_i$, when all the overheads of scheduling and resource claiming are included under maximum speed $S_{max}$, actual execution time $a_i$ which is the actual time taken by the task during execution, and skip factor $sf_i$ which specify the task's tolerance to deadline misses. That means that the distance between two consecutive skips must be at least $sf_i$ periods. By definition, the actual execution time of any task is always less than or equal to its worst-case execution time, that is, $a_i \leq c_i$. We further assume that the relative deadline $d_i$ is equal to the period $p_i$. $T_{i,j}$ denotes the $j^{th}$ instance or job of task $T_i$. Every instance of a task is either red or blue [4]. A red task instance must complete before its deadline: A blue task instance can be aborted at any time. However, if a blue instance completes successfully, the next task instance is still blue. We use the term hyper-period *P* to refer to the least common multiple of all task periods, that is

$$P = LCM (p_1, p_2, …, pn)$$

We assume pre-emptive scheduling, and that the pre-emption and speed change overheads can be incorporated in $c_i$ if necessary. The process descriptor of $T_i$ is augmented to include two fields related to the CPU speed: a nominal speed $S_i^{nom}$, which is the default speed assigned to the task when it is about to be dispatched, and an actual speed $S_i$, which is the speed at which the task is being executed at the specific time instant. Under a constant speed $S$, the execution time of task $T_i$ is $c_i/S$. The utilization of task $T_i$ under CPU speed $S$ is given by $u_i(S) = c_i/(p_i S)$. The aggregate utilization of the task set (under maximum speed) is given by

$$U_{tot} = \sum_{i=1}^{n} c_i/p_i$$

In this paper, we assume that the execution time scales linearly with the CPU speed, ignoring memory stall effects. This is a conservative but safe assumption since it overestimates the new execution time when the CPU speed is reduced [11]. Hence, it does not affect the schedulability analysis.

With the above system models, our problem can be formulated as follows:

Given system $\Gamma = \{T_1, T_2, \cdots, T_n\}$, $T_i = (p_i, c_i, a_i, sf_i)$, $i = 0, \cdots, n$, schedule $\Gamma$ with a dynamic scheduling algorithm on a variable voltage processor with discrete supply voltage levels $V = \{V_{min}, ..., V_{max}\}$ and corresponding processor speeds $S = \{S_{min}, ..., S_{max}\}$ such that all constraints with a skip factor $sf$ are guaranteed and the energy consumption is minimized.

## 3. EXISTING ALGORITHMS

In this paper, existing three scheduling algorithms designed for overloaded real-time systems that allow skips are considered for energy efficiency. First two scheduling algorithms namely Red Tasks Only (RTO) algorithm and Blue When Possible (BWP) algorithm were introduced by Koren and Shasha [4]. In RTO algorithm, red instances are scheduled as soon as possible according to Earliest Deadline First (EDF) algorithm while blue ones are always rejected. The BWP is an improvement of RTO, and this schedules blue instances whenever their execution does not prevent the red ones from completing within their deadlines. In other words, blue instances are served in background relatively to red instances. The third algorithm Red as Late as Possible (RLP) algorithm [2] brings forward the execution of blue task instances so as to minimize the ratio of aborted blue instances, thus enhancing the QoS (i.e., the total number of task completions) of periodic tasks. It considered two factors,

- If there are no blue task instances in the system, red task instances are scheduled as soon as possible according to the EDF algorithm.
- If blue task instances are present in the system, they are scheduled as soon as possible according to the EDF algorithm, while red task instances are processed as late as possible according to the EDL algorithm.

### 3.1. RTO Algorithm

In this algorithm red instances are scheduled as soon as possible according to EDF algorithm, while blue ones are always rejected. Deadline ties are broken in favour of the task with the earliest release time. In the deeply red model where all tasks are synchronously activated and the first $sf_i-1$ instances of every task $T_i$ are red, this algorithm is optimal. Initial RTO schedule is illustrated in Figure 1 using the task set T = $\{T_0, T_1, T_2, T_3, T_4\}$ of five periodic tasks whose parameters are described in Table 1. Tasks have uniform skip parameter $sf_i = 2$ and the total processor utilization factor $U_{tot}$ is equal to 1.15.

Table 1: Task Parameters

| Task | $T_1$ | $T_2$ | $T_3$ | $T_4$ | $T_5$ |
|------|-------|-------|-------|-------|-------|
| $c_i$ | 3 | 4 | 1 | 7 | 2 |
| $p_i$ | 30 | 20 | 15 | 12 | 10 |

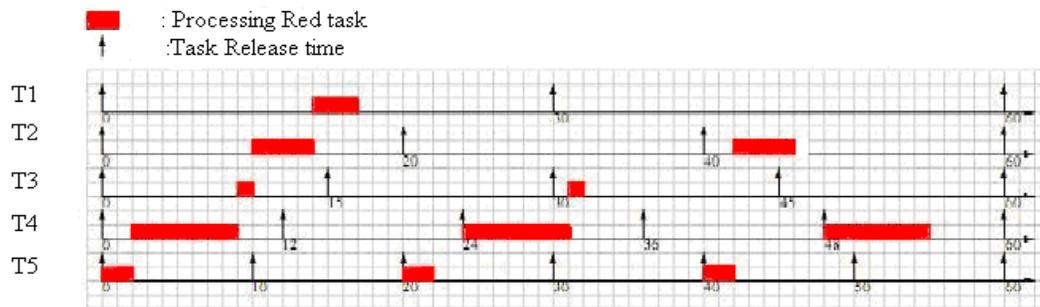

Figure 1: RTO Schedule ($sf_i = 2$)

As we can see, the distance between every two skips is exactly $sf_i$ periods, thus offering only the minimal guaranteed QoS level for periodic tasks.

This RTO algorithm is implemented by creating two queues. One is the red queue and the other one is the blue queue. In the beginning the instances of all the tasks are created. These task instances may be either red or blue. The red task instances are queued in the red queue and the blue task instances are queued in the blue queue. The task instances in both the queues are sorted in the order of increasing deadline.

The red instance with least deadline will be executed first. The scheduler places the task instances generated periodically in the appropriate queue for execution. The red instances alone are executed. The blue instances are left as it is without executing. The red hit variable is incremented after each successful completion of the red task instances. The red miss variable is incremented if a red task instance misses the deadline. If the distance between two skips is less than the skip factor then it is considered as a miss. The blue miss variable is incremented after each completion of the period since there will not be any execution of blue tasks instances. Whenever a blue task instance is generated it will be missed definitely.

The RTO scheduler creates a feasible schedule for the hyper-period of a given task set. Then the success ratio is calculated for the task set under schedule. The success ratio is ratio of the total number of hits to the total number of task instances generated in a hyper-period.

### 3.2. BWP Algorithm

This algorithm schedules blue instances whenever their execution does not prevent the red ones from completing within their deadlines. In that sense, it operates in a more flexible way. Deadline ties are still broken in favour of the task with the earliest release time. Figure 2 shows an illustrative example of BWP scheduling using the task set previously mentioned in Table 1.

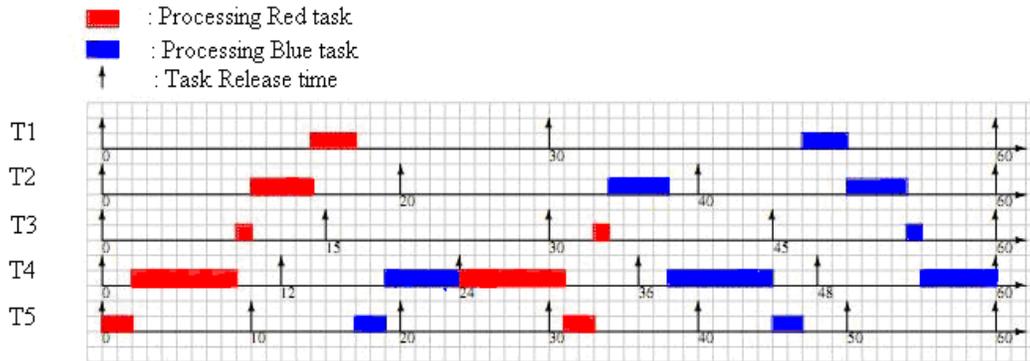

Figure 2: BWP schedule ($sf_i = 2$)

This Blue When Possible (BWP) algorithm is implemented by creating two queues. One is the red queue and the other one is the blue queue. In the beginning the instances of all the tasks are created. These task instances may be either red or blue. The red task instances are queued in the red queue and the blue task instances are queued in the blue queue. The task instances in both the queues are sorted in the order of increasing deadline.

The red instance with least deadline will be executed first. After the completion of the first red instance the next red instance with least deadline will be executed. When there are no red task instances to execute in the red queue, then the blue task instance with the least deadline is executed. If a red task instance is created, then immediately this executing blue task will abort its execution and the red task instance will be executed. As soon as the period of each task is completed, a new task instance is created and queued in the appropriate queue for execution.

The red hit variable is incremented after each successful completion of the red task instances. The red miss variable is incremented if a red task instance misses the deadline. If the distance between two skips is less than the skip factor then it is considered as a skip. The blue hit variable is incremented after each successful completion of the blue task instances. The blue miss variable is incremented if a blue task instance misses the deadline or if it is aborted.

### 3.3. RLP Algorithm

The main drawback of BWP relies on the fact that blue task instances are executed as background tasks. This leads to abort partially or almost completely executed blue task instances, thus wasting processor time.

The objective of RLP algorithm is to bring forward the execution of blue task instances so as to minimize the ratio of aborted blue instances, thus enhancing the actual QoS (i.e., the total number of task completions) of periodic tasks. From this perspective, RLP scheduling algorithm, which is a dynamic scheduling algorithm, is specified by the following behaviour:

- If there are no blue task instances in the system, red task instances are scheduled as soon as possible according to the EDF (Earliest Deadline First) algorithm.
- If blue task instances are present in the system, these ones are scheduled as soon as possible according to the EDF algorithm (note that it could be according to any other heuristic), while red task instances are processed as late as possible according to the EDL algorithm.

Deadline ties are always broken in favour of the task with the earliest release time. The main idea of this approach is to take advantage of the slack of red periodic task instances.

Determination of the latest start time for every red request of the periodic task set requires preliminary construction of the schedule by a variant of the EDL algorithm taking skips into account [12]. In the EDL schedule established at time τ, we assume that the instance following immediately a blue instance which is part of the current periodic instance set at time τ, is red. Indeed, none of the blue task instances is guaranteed to complete within its deadline. Moreover, Silly-Chetto in [12] proved that the online computation of the slack time is required only at time instants corresponding to the arrival of a request while no other is already present on the machine.

In our case, the EDL sequence is constructed not only when a blue task is released (and no other was already present) but also after a blue task completion if blue tasks remain in the system (the next task instance of the completed blue task has then to be considered as a blue one). Note that blue tasks are executed in the idle times computed by EDL and are of same importance beside red tasks (contrary to BWP which always assigns higher priority to red tasks).

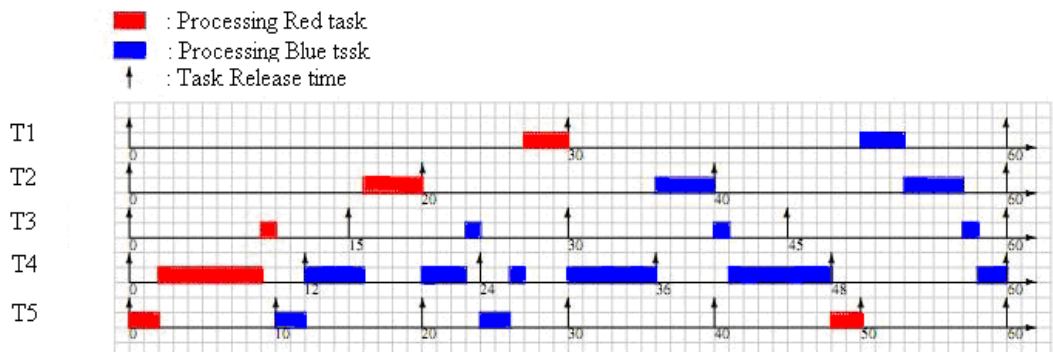

Figure 3: RLP Schedule ($sf_i = 2$)

## 4. PROPOSED ENERGY EFFICIENT SCHEDULING TECHNIQUES

A major trend in the microprocessor industry is towards energy-efficient mobile computing for maximal battery life using the concept of performance on demand. The basic idea is to run the CPU at a voltage and frequency that satisfies the current performance requirement. Dynamic voltage and frequency scaling is a very effective technique for reducing CPU energy [13] [14]. Significant energy benefits can be achieved by recognizing that peak performance is not always required and therefore the operating voltage and frequency of the processor can be dynamically adapted based on instantaneous processing requirement. Examples include the Intel Pentium III SpeedStep technology [15] which lets the user run the processor at a lower voltage and frequency when using the battery, LongRun technology from Transmeta's Crusoe [16] and PowerNOW! Technology from AMD [17]. In this paper, we propose energy-efficient technique to the above existing real-time scheduling algorithms that can exploit the variable voltage and frequency hooks available on processors for improving energy efficiency.

### 4.1. DVS with Processor Reclamation

DVS allows adjusting processor voltage and frequency at runtime. DVS can be implemented at various levels of a system, such as in the processor, in the OS scheduler, in the compiler or in the application. Operating system is the only component with an overview of the entire system, including task constraints and status, resource usage, etc. Therefore, the most effective and efficient approaches to reduce energy consumption can be achieved with proper task scheduling algorithms. It is time consuming to find an optimal schedule where energy consumption is minimized and all timing constraints are met. Many previous works either proposed offline scheduling for large energy reduction, or used heuristic methods to reduce scheduling overhead.

However, while the former approaches are inflexible and too costly to store in memory, the latter ones may not realize the full potential of energy savings.

Our approach involves pre-computing a global nominal speed for all the tasks statically, and applying the dynamic reclamation/speed adjustment techniques online whenever possible. At dispatch time, the speed of each job of $T_i$ is first set to the nominal speed $S^{nom}$. However, its actual speed $S_i$ may be even lower after the application of the dynamic slack reclamation techniques. The nominal speed must be carefully chosen to guarantee the deadlines of the mandatory jobs. At run-time, it is possible to further reduce the actual CPU speed, and consequently reduce the energy consumption, by observing that the schedule has idle intervals due to the optional jobs that are skipped. Thus, it is possible to use this slack time for dynamic slow-down making it possible to improve the energy efficiency of the system.

If, at run-time, the mandatory jobs complete execution before their worst case execution time, then it is possible to exploit the unused processor time to further minimize the energy consumption by performing dynamic speed reduction. We perform dynamic speed slow-down by using the Dynamic Reclaiming Algorithm (DRA) [8]. DRA detects early task completions by comparing the actual schedule to the static optimal schedule. In this schedule, all the jobs run at the same speed, namely the nominal speed $S^{nom}$, through which all the (selected) jobs will be able to meet their deadlines even under a worst-case workload. DRA determines the amount of processor time that a dispatched job can safely use to slow down its speed. This additional processor time is used to calculate the new lower speed of a currently dispatched job. A main feature of the scheme is to calculate this additional time quickly, and without affecting the feasibility of already selected tasks. The earliness is computed in such a way to allow the low-priority tasks to use the slack time of completed high priority tasks. The exact formula for calculating the earliness and determining the reduced speed, as well as the details of the DRA is discussed in [8].

### 4.2. Hybrid Approach

Extensive power aware scheduling techniques have been published for energy reduction, but most of them have been focused solely on reducing the processor energy consumption. While the processor is one of the major power hungry units in the system, other peripherals such as network interface card, memory banks, disks also consume significant amount of power. Dynamic Power Down (DPD) technique is used to shut down a processing unit and peripheral devices to save power when it is idle. There is a minimal time interval that the device can be feasibly shut down with positive energy-saving gain. Only when the slack time or idle time is more than this minimum time interval the DPD is used.

In this paper, we combined DPD and DVS techniques in order to make the above existing algorithms energy-efficient with less overhead and without affecting the QoS provided by these algorithms to the system. First DVS is applied, and then if possible DPD is applied to further reduce the energy.

### 4.2. Overall Architecture

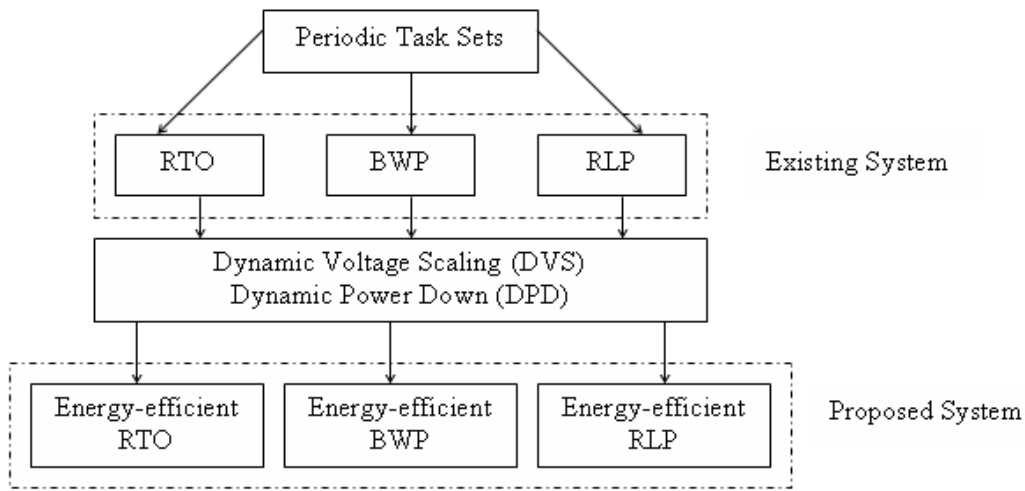

Figure 4. Overall Architecture Block Diagram

The architecture block diagram of the proposed system is shown in Figure 4. Randomly generated task sets are given as input for the three existing weakly-hard real-time scheduling algorithms RTO, BWP and RLP. The existing three algorithms are made energy efficient using the above discussed DVS and DPD techniques. These two techniques are applied to the algorithms in a required time multiplexed manner so that it involves less overhead and reduces the energy effectively without affecting the QoS provided by the algorithms to the System.

## 5. SIMULATION AND ANALYSIS OF RESULTS

### 5.1. Simulation

The input to the algorithms is the task set with parameters as mentioned in the task model. These parameters are generated randomly. The period of a task is randomly generated within a range 3 to 100, to maintain the hyper period which is the Least Common Multiple (LCM) of the periods for the given task set within a limit. The periods of the second, third and fourth tasks are generated as random multiples of the first one. The parameter computation time is selected in the range 1 to 15, and it also maintained to be less than or equal to period, as the computation time is mostly less than the period in the case of real-time system. When each task repeats itself, a red instance or a blue instance of the task should be created. This instance creation is made random.

The simulated algorithms generate the average success ratio by calculating the success ratio of each task and calculating the average of them. The algorithms are first tested with two numbers of tasks. The average success ratio for two tasks with a simulation test running ten times are noted separately for all the three algorithms This action is repeated for  the increased number of tasks. Each point in the graph is an average of ten simulation runs

The same method is followed for the energy calculation in the simulated energy efficient algorithms also. The maximum number of tasks taken for the simulation is 10 tasks as the number of independent tasks in a task set of a real-time system may not exceed this number in general.

## 5.2. Result Analysis

The existing three scheduling algorithms for the weakly-hard real-time systems are simulated and the graph is plotted with the values of success ratio of periodic task set against the number of tasks given as input as shown in Figure 5. From the figure it is noted that RLP gives the best QoS out the three existing algorithms. The RTO algorithm schedules only red task instances. This algorithm never even tries to schedule any blue task instance. So, it is clear from the graph that it provides only less QoS than the other two. Even though both BWP and RLP algorithms schedule, both red and blue task instances, RLP gives the maximum QoS among the three.

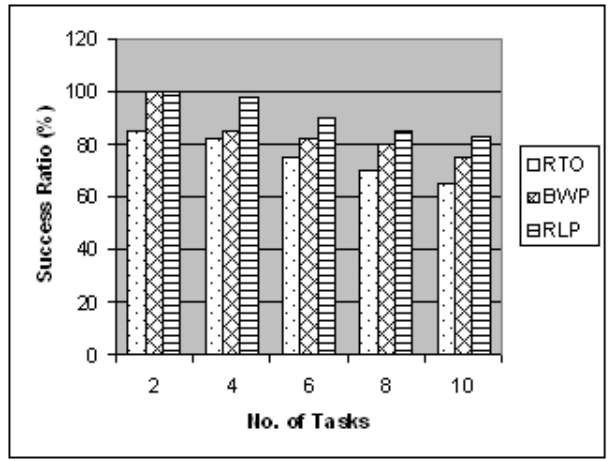

Figure 5. Success Ratio

The above existing algorithms are simulated with energy efficiency and the graph is plotted with the values of normalized energy consumption against the number of tasks in the task set generated. This is given in Figure 6, and shows that out of three algorithms, RTO consumes lesser energy. This is due to the fact that RTO schedules only the red task instance leaving all blue task instances. So, more slack time is available for applying DVS and DPD techniques in appropriate slack periods thereby reducing the energy consumption to a maximum level. Even though both BWP and RLP algorithms schedule, both red and blue task instances, RLP consumes lesser energy.

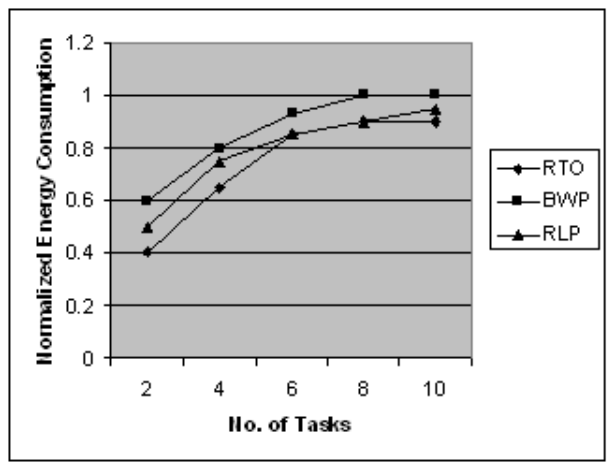

Figure 6. Normalized Energy Consumption

International Journal of Computer Science & Information Technology (IJCSIT), Vol 2, No 6, December 2010

## 6. CONCLUSION

Three existing scheduling algorithms for weakly-hard real-time systems were simulated for energy efficiency without affecting the QoS using the proposed hybrid approach. From the simulation results it is found that the RLP scheduling algorithm gives better QoS to the system and the RTO algorithm consumes less energy among the three. Our aim is to reduce the energy consumption while maintaining the same QoS offered by the algorithms. Even though RTO algorithm consumes less energy among the three, it never schedules the blue instances of task set in the system, which decreases the QoS of the system. However scheduling of blue task instances in addition to mandatory red task instances will increase the QoS of the system. It is also studied from the results that, when QoS as well as energy efficiency are to be maintained together, the RLP scheduling algorithm is the efficient one. The simulation results show that RLP algorithm is more efficient in considering both QoS and energy consumption among the three algorithms studied.

The proposed energy-efficient algorithms are designed for single processor system which considers only independent tasks. However, there exists scope for it being extended to multi processor or distributed systems which consider dependant tasks with precedence and resource constraints along with the timing constraints. Though such an extension requires more efficient algorithms and power scheduling techniques, it can be worked out.

[15]     http://www.intel.com/mobile/pentiumIII/ist.htm
[16]     http://www.transmeta.com/crusoe/lowpower/longrun.html
[17]     http://www.amd.com/products/cpg/mobile/powernow.html

**Authors**

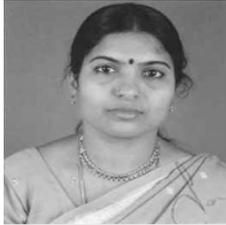

**Mrs. Santhi Baskaran** received her B.E. degree in Computer Science and Engineering from University of Madras, Chennai, India in 1989 and M.Tech. degree in Computer Science and Engineering from Pondicherry University, Puducherry, India in 1998. She served as Senior Lecturer and Head of the Computer Technology Department, in the Polytechnic Colleges, Puducherry, India for eleven years, since 1989. She joined Pondicherry Engineering College, Puducherry, India in 2000 and currently working as Associate Professor in the Department of information Technology. Now she is pursuing her PhD degree in Computer Science and Engineering. Her areas of interest include Real-time systems, embedded systems and operating systems. She has published research papers in International and National Conferences. She is a Life member of Indian Society for Technical Education and Computer Society of India.

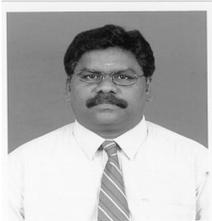

**Prof. Dr. P. Thambidurai** is a Member of IEEE Computer Society. He received his PhD degree in Computer science from the Alagappa University, Karaikudi, India in 1995. From 1999, he served as Professor and Head of the Department of Computer Science & Engineering and Information Technology, Pondicherry Engineering College, Puducherry, India, till August 2006. Now he is the Principal for Perunthalaivar Kamarajar Institute of Engineering and Technology (PKIET) an Government institute at Karaikal, India. His areas of interest include Natural Language Processing, Data Compression and Real-time systems. He has published over 50 research papers in International Journals and Conferences. He is a Fellow of Institution of Engineers (India). He is a Life member of Indian Society for Technical Education and Computer Society of India. He served as Chairman of Computer Society of India, Pondicherry Chapter for two years. Prof. P.Thambidurai is serving as an Expert member to All India Council for Technical Education (AICTE) and an Adviser to Union Public Service Commission (UPSC), Govt. of India. He is also an Expert Member of IT Task Force and Implementation of e-Governance in the UT of Puducherry.